\documentclass[11pt,draftcls, onecolumn]{IEEEtran}
\ifCLASSINFOpdf
\else
\fi
\usepackage{bm,amsmath,mathrsfs,amsfonts,graphicx}
\usepackage{algpseudocode,cite}
\usepackage{algorithm}

\begin{document}

\newcommand{\bGamma}{\bm{\Gamma}}
\newcommand{\bSigma}{\bm{\Sigma}}
\newcommand{\bOmega}{\bm{\Omega}}

\newcommand{\bphi}{\bm{\phi}}
\newcommand{\bpi}{\bm{\pi}}
\newcommand{\bdelta}{\bm{\delta}}
\newcommand{\bomega}{\bm{\omega}}
\newcommand{\bgamma}{\bm{\gamma}}
\newcommand{\bepsilon}{\bm{\epsilon}}
\newcommand{\blambda}{\bm{\lambda}}
\newcommand{\btheta}{\bm{\theta}}
\newcommand{\bpsi}{\bm{\psi}}
\newcommand{\bmeta}{\bm{\eta}}

\newcommand{\tilDelta}{\tilde{\Delta}}
\newcommand{\tilTheta}{\tilde{\Theta}}
\newcommand{\tiltheta}{\tilde{\theta}}

\newcommand{\bA}{\mathbf{A}}
\newcommand{\bC}{\mathbf{C}}
\newcommand{\bF}{\mathbf{F}}
\newcommand{\bG}{\mathbf{G}}
\newcommand{\bH}{\mathbf{H}}
\newcommand{\bI}{\mathbf{I}}
\newcommand{\bJ}{\mathbf{J}}
\newcommand{\bM}{\mathbf{M}}
\newcommand{\bN}{\mathbf{N}}
\newcommand{\bP}{\mathbf{P}}
\newcommand{\bQ}{\mathbf{Q}}
\newcommand{\bR}{\mathbf{R}}
\newcommand{\bS}{\mathbf{S}}
\newcommand{\bT}{\mathbf{T}}
\newcommand{\bU}{\mathbf{U}}
\newcommand{\bV}{\mathbf{V}}
\newcommand{\bW}{\mathbf{W}}
\newcommand{\bX}{\mathbf{X}}
\newcommand{\bZ}{\mathbf{Z}}

\newcommand{\ba}{\mathbf{a}}
\newcommand{\bb}{\mathbf{b}}
\newcommand{\bd}{\mathbf{d}}
\newcommand{\be}{\mathbf{e}}
\newcommand{\mbf}{\mathbf{f}}
\newcommand{\bg}{\mathbf{g}}
\newcommand{\bh}{\mathbf{h}}
\newcommand{\bn}{\mathbf{n}}
\newcommand{\bp}{\mathbf{p}}
\newcommand{\bq}{\mathbf{q}}
\newcommand{\br}{\mathbf{r}}
\newcommand{\bs}{\mathbf{s}}
\newcommand{\bu}{\mathbf{u}}
\newcommand{\bv}{\mathbf{v}}
\newcommand{\bw}{\mathbf{w}}
\newcommand{\bx}{\mathbf{x}}
\newcommand{\by}{\mathbf{y}}
\newcommand{\bz}{\mathbf{z}}

\newcommand{\hbeta}{\hat{\beta}}
\newcommand{\htheta}{\hat{\theta}}
\newcommand{\hsigma}{\hat{\sigma}}

\newcommand{\hp}{\hat{p}}
\newcommand{\hn}{\hat{n}}
\newcommand{\hr}{\hat{r}}
\newcommand{\hs}{\hat{s}}
\newcommand{\hx}{\hat{x}}

\newcommand{\hN}{\hat{N}}

\newcommand{\hbSigma}{\hat{\bm{\Sigma}}}

\newcommand{\hba}{\hat{\mathbf{a}}}
\newcommand{\hbs}{\hat{\mathbf{s}}}
\newcommand{\hbv}{\hat{\mathbf{v}}}

\newcommand{\hbP}{\hat{\mathbf{P}}}
\newcommand{\hbR}{\hat{\mathbf{R}}}
\newcommand{\hbW}{\hat{\mathbf{W}}}

\newcommand{\dif}{\text{d}}

\newcommand{\bbC}{\mathbb{C}}
\newcommand{\bbR}{\mathbb{R}}
\newcommand{\bbN}{\mathbb{N}}
\newcommand{\bbZ}{\mathbb{Z}}

\newcommand{\calA}{\mathcal{A}}
\newcommand{\calB}{\mathcal{B}}
\newcommand{\calC}{\mathcal{C}}
\newcommand{\calD}{\mathcal{D}}
\newcommand{\calE}{\mathcal{E}}
\newcommand{\calF}{\mathcal{F}}
\newcommand{\calH}{\mathcal{H}}
\newcommand{\calI}{\mathcal{I}}
\newcommand{\calN}{\mathcal{N}}
\newcommand{\calM}{\mathcal{M}}
\newcommand{\calS}{\mathcal{S}}
\newcommand{\calT}{\mathcal{T}}
\newcommand{\calV}{\mathcal{V}}
\newcommand{\calW}{\mathcal{W}}
\newcommand{\calX}{\mathcal{X}}
\newcommand{\calY}{\mathcal{Y}}

\newcommand{\tlA}{\tilde{A}}

\newcommand{\tlp}{\tilde{p}}
\newcommand{\tls}{\tilde{s}}
\newcommand{\tlv}{\tilde{v}}

\newcommand{\tlcalI}{\tilde{\calI}}

\newcommand{\barn}{\bar{n}}
\newcommand{\barr}{\bar{r}}
\newcommand{\bary}{\bar{y}}

\newcommand{\barS}{\bar{S}}
\newcommand{\barX}{\bar{X}}
\newcommand{\barY}{\bar{Y}}

\newcommand{\barba}{\bar{\ba}}
\newcommand{\barby}{\bar{\by}}
\newcommand{\barbz}{\bar{\bz}}

\newcommand{\tlbA}{\tilde{\bA}}
\newcommand{\tlbR}{\tilde{\bR}}

\newcommand{\tlbp}{\tilde{\bp}}
\newcommand{\tlbs}{\tilde{\bs}}
\newcommand{\tlbv}{\tilde{\bv}}

\newcommand{\tc}{\text{c}}
\newcommand{\td}{{\text{d}}}

\newcommand{\suml}{\sum\limits}
\newcommand{\prodl}{\prod\limits}
\newcommand{\minl}{\min\limits}
\newcommand{\maxl}{\max\limits}
\newcommand{\infl}{\inf\limits}
\newcommand{\supl}{\sup\limits}
\newcommand{\liml}{\lim\limits}
\newcommand{\intl}{\int\limits}
\newcommand{\bigcupl}{\bigcup\limits}

\newcommand{\opconv}{\text{conv}}

\newcommand{\eref}[1]{(\ref{#1})}

\newcommand{\sinc}{\text{sinc}}
\newcommand{\tr}{\text{Tr}}
\newcommand{\var}{\text{Var}}
\newcommand{\cov}{\text{Cov}}
\newcommand{\tth}{\text{th}}

\newenvironment{vect}{\left[\begin{array}{c}}{\end{array}\right]}
\newtheorem{theorem}{Theorem}
\newtheorem{lemma}{Lemma}

%
\title{A Novel Sparsity-Based Approach to Recursive Estimation of Dynamic Parameter Sets}
%
%
%

\author{Ashkan~Panahi,~\IEEEmembership{Member,~Student Member,}
        Mats~Viberg,~\IEEEmembership{Fellow,~IEEE,}
\thanks{A. Panahi and M. Viberg are with the Department
of Signals and Systems, Chalmers University, Gothenburg,
Sweden e-mail: \{ashkanp, Viberg\}@chalmers.se .}
\thanks{This work is supported by the Swedish Research Council (VR).}
\thanks{Manuscript received April 19, 2005; revised January 11, 2007.}}

%
%

\markboth{Journal of \LaTeX\ Class Files,~Vol.~6, No.~1, January~2007}%
{Shell \MakeLowercase{\textit{et al.}}: Bare Demo of IEEEtran.cls for Journals}
%



\maketitle

\begin{abstract}
We consider the problem of estimating a variable number of parameters with a dynamic nature. A familiar example is finding the position of moving targets using sensor array observations.  The problem is challenging in cases where either the observations are not reliable or the parameters evolve rapidly.  Inspired by the sparsity based techniques, we introduce a novel Bayesian model for the problems of interest and study its associated recursive Bayesian filter.  We propose an algorithm approximating the Bayesian filter, maintaining a reasonable amount of calculations. We compare by numerical evaluation the resulting technique to state-of-the-art algorithms in different scenarios. In a scenario with a low SNR, the proposed method outperforms other complex techniques. 
\end{abstract}


%
\IEEEpeerreviewmaketitle

\section{Introduction}\label{sec:intro}
\IEEEPARstart{E}{stimating} a dynamic set of parameters is a highly useful and  wide area of research, with a long and fruitful history \cite{chen2003bayesian}. Indeed, noticing the ever increasing application of the Kalman filter and its variants to many newly developed technologies is enough to understand the importance of this topic. In this context, the quest for modified techniques usually concerns cases where either the currently existing methods fail to meet the computational limitations, or result in an insufficient precision. The latter may also be either  due to an inconsistent model, on which the technique is based, or simply because of improper approximations. From this perspective, one finds certain estimation problems, for example the ones concerning data generated by a sensor array, more challenging. The reason is that the associated models, being capable of capturing the desired properties of the parameters, are so complicated that standard design methods by them lead to computationally intractable techniques. Thus, appealing to proper approximations is inevitable  in those cases. This paper addresses these problems and aims to provide a modified approximate estimation technique. The emphasis here is on maintaining a low computational complexity, while maintaining the statistical properties of the estimates.

The central idea in estimating a time varying parameter is that a parameter following a well-structured temporal model has locally  correlated samples. Thus, they can be fused to improve the quality of estimation for a specific sample. This is particularly known as parameter filtering \cite{jazwinski2007stochastic}.
The basic ideas of filtering can be easily observed in the pioneering studies of Wiener, initiating the field of adaptive filtering \cite{kailath1981lectures}. Later, the seminal work of Kalman framed adaptive filtering  into a rigorous statistical context, and showed a case, where statistically efficient estimates could be exactly calculated by a recursive method \cite{kalman1960new}. Soon after Kalman, Ho and Lee generalized this idea to the so called Markov Chain (MC) models, comprising of parameter evolution and measurement models \cite{ho1964bayesian}. Their solution is generally called Recursive Bayesian Filtering (RBF). The main advantage of the RBF is that it is highly adaptive to different application specifications, including a non-stationary behavior \cite{anderson2012optimal}. However, it requires storing and integrating posterior densities. Approximate techniques such as the Extended Kalman Filter (EKF) \cite{haykin2001kalman,ljung1979asymptotic} and Unscented Kalman Filter (UKF) \cite{julier1997newUKF} are commonly used to perform this. Due to their local nature, they perform poorly, when multi-modal distributions are considered.  The advent of statistical sampling and Monte Carlo methods provided an alternative method of implementing recursive Bayesian filters, by the so called Markov Chain Monte Carlo (MCMC) method. The resulting filter is generally known as the particle filter \cite{kitagawa1996monte,carpenter1999improved,van2000unscented}.

The difficulty arises in applying the above to problems such as radar detection, where the data is generated by a sensor array. This is due to multiple reasons, discussed in the sequel. The first reason is that a MC model is not directly applicable. To elaborate on this, note that the corresponding measurement model for data generated by a sensor array consists of two distinct set of parameters, known as amplitude and position parameters. In many applications, the amplitudes evolve rapidly in time, resulting in highly uncorrelated samples. Thus, only in the sense of position parameters one may perceive a Bayesian filter. To remain in the realm of RBF, it is still necessary to handle the amplitudes in a Bayesian manner. The second reason is that the observation model of the applications of interest is nonlinear, and estimation through them usually leads to the local minima problem. In the same manner, nonlinearity results in posterior multimodality, which not only complicates estimation, but also makes the posterior calculation difficult. The third reason is related to the fact that the time evolution model of the position parameters concerns varying order. Take the radar example. In the course of observation, it is perceivable that some targets may be introduced or removed from the observation scene. In a more elaborate model, a single target may spawn multiple future targets. A MC model capturing the dynamics of such a system is complex and its corresponding sequential Bayesian filter can only be derived in an abstract form. To reduce the computational cost without introducing too much error, this filter needs to be approximately parametrized. This is generally a challenging task.
\subsection{Literature Survey}
Due to the above, one may find different approaches in the literature to recursive filtering of the sensor array data. According to different representations of the problem of interest, these methods are developed under different names. More specifically, the parametric (Kalman filter-based), spectral-based  and subspace-based representations give rise to filtering techniques under similar titles. Some spectral based techniques can be found, e.g in \cite{lototsky1997nonlinear,heidenreich2009morphological,rankine2007if}.  
The subspace tracking approaches have also been recently studied and applied in the literature \cite{moonen1992singular,stewart1992updating,yang1995projection}. The semi-parametric sparsity-based techniques are also rapidly emerging in literature under the title of sparsity tracking \cite{mecklenbrauker2013sequential,vaswani2008kalman,lustig2006kt}. The filtering techniques can be also categorized from a different perspective.  Many studies consider a case where preliminary parameter estimates are provided, relying only on their corresponding data. This is called target tracking and is favorable in occasions, such as some radar detection problems, where only the preliminary estimates are accessible for process \cite{bar2011tracking, blackman1986multiple,lipton1998moving}. In contrast to target tracking, the recent attempts to directly use sensor data to perform parameter filtering is often referred to as Track-Before-Detect (TBD), but this is not a generic term \cite{davey2008comparison, tonissen1996peformance}. 

The above techniques deal with the aforementioned difficulties in different ways. The target tracking and the subspace tracking techniques  do not suffer from lack of amplitude models, while other TBD approaches either assume a specific amplitude model, depending on the application or eliminate them by assuming a Bayesian model and integration \cite{tonissen1998maximum}. The amplitude models usually involve hyper-parameters, for which simple time evolution models are considered. The parametric formulation is the most precise likelihood based approach \cite{tonissen1998maximum}, but is numerically sensitive to nonlinearity. The Joint Probabilistic Data Association (JPDA) and Probabilistic MultiHypothesis Tracker PMHT \cite{willett2002pmht} methods are  popular examples of parametric target tracking \cite{bar1975tracking,fortmann1983sonar}. Instead, the methods leading to spectral estimation such as subspace-based and sparsity-based techniques trade off precision in favor of numerical stability.  Moreover, particle filtering is nowadays a common approach to overcome multi-modality \cite{arulampalam2002tutorial}. Concerning the issue with variable order, many related studies consider a fairly general model, where the parameters have a fixed probability to survive, disappear or appear at the next time instant. In the recent literature, this is formulated as a Random Finite Set (RFS) model, also considered here and referred to as the standard model \cite{vo2005sequential}. The RFS based representation not only provides a formal definition of the time evolution model, but also suggests certain approximation techniques. For example the Probability Hypothesis Density (PHD) filter provides a method to overcome the so-called data association problem in target tracking through approximating the RFS-based posteriors by a Poisson process \cite{mahler2000theoretical}. The data association problem is due to the fact that the preliminary estimates are not generally labeled by their corresponding true parameter. More elaborate examples of such can be found in \cite{svensson2011set,crouse2011set}.   

\subsection{Motivation}

In the problems of interest herein, the RBF approach needs to be approximated and the performance of all the techniques in the prequel is limited by the precision of their underlying approximation. From this perspective, these techniques can be divided into three groups: The locality based approaches such as EKF and UKF, the ones based on stochastic sampling, i.e. particle filters, and other model-based approximations such as the ones in the PHD filtering. The latter is normally based on minimizing the Kullback-Leibler (KL) distance between the resulting posteriors and a parametrized model set, which is applicable only  if the minimization has a tractable solution. Clearly, the choice of approximation depends on the type of filter. For example, a locality based approximation is not appropriate for parametric filtering, where multiple local minima are present. In general, particle filters are always applicable, but need a higher computational effort (number of particles) than the other techniques to provide the desired precision. The precision of the methods such as the PHD filter depends on how well the approximate model fits to the exact one. Practically speaking, this restricts such methods to a high SNR or a slowly varying case. Moreover, the target tracking performance is also dependent on the quality of the preliminary estimates, which considerably limits the SNR range of application for these techniques.

In this paper, we study a different opportunity provided by the findings in the field of sparsity-based estimation, especially the Least Absolute Shrinkage and Selection Operator (LASSO) \cite{tibshirani,basis,stoica2011spice}. Recently, the inspiring work of Stoica et al in \cite{stoica2011spice,stoica2012spice} has provided an important Bayesian insight into this approach, which we slightly modify here to fit the RFS framework. Using this model for observation and considering the standard RFS based time evolution model, we investigate on the resulting RBF. The RBF is again intractable and needs approximation. On the other hand, it is observed that the convexity of LASSO yields to unimodality of the posterior distributions. Thus, it is favorable to use local approximations, similar to EKF. We develop a local expansion technique performed on the abstract space of finite sets and apply it to the proposed RFS, leading to a tractable filter.

\subsection{Mathematical Notation}
In this paper, $\bbR$, $\bbR_+$ and $\bbC$ refer to the set of real, non-negative real and complex numbers, respectively. The notation $\tr(\ldotp)$ denotes the trace operator and $|.|$ shows either the absolute (in the case of a numerical argument), or the cardinality (in the case of a set argument) of the argument. Moreover, $(\ldotp)_+$ denotes the positive part of its real argument. We also define  an assignment $R$ between finite sets $A$ and $B$ as a subset of $A\times B$ satisfying the following conditions
\begin{itemize}
\item $\forall (a_1,b_1)\in R,(a_2,b_2)\in R;\quad a_1=a_2\rightarrow b_1=b_2 $
\item $\forall (a_1,b_1)\in R,(a_2,b_2)\in R;\quad b_1=b_2\rightarrow a_1=a_2 $
\end{itemize}
Moreover, we define the domain sets of $R$ as the elements in $A$ and $B$, included in $R$, i.e.
\begin{itemize}
\item $d_1(R)=\{a\in A\mid \exists b\in B,\ (a,b)\in R\}$ 
\item $d_2(R)=\{b\in B\mid\exists a\in A,\  (a,b)\in R\}$
\end{itemize}
Throughout the paper,  $+$ and $-$ subscripts or superscripts denote parameter values right after and before an observation, respectively. The primed parameters are usually related to the ones at a previous time instant. The notation $p(\ldotp)$ denotes the probability density function (pdf) of its argument and $Q(\ldotp,\ldotp)$ represents the transitional probability between consecutive samples. 
%

\section{Problem Formulation}\label{sec:problem}
\subsection{Observation Model}
Consider a compact subset $\Theta\subset\bbR$ and a smooth basis manifold $\ba :\Theta\to \bbC^m$. Further, consider a vector data set $\{\bx(t)\in\bbC^m\}_{t=1}^\infty$, observed through the following model:
\begin{equation}\label{eq:model}
\bx(t)=\suml_{k=1}^{n_t}\ba (\theta_k(t))s_k(t)+\bn (t)
\end{equation}  
where $t$ is the time index, the sets $\{\theta_k(t)\in\Theta\}$ and $\{s_k(t)\in\bbC\}$ are called position and amplitude parameters, respectively and $\{\bn(t)\}$ denotes the additive noise, assumed to be a centered Gaussian, white and stationary process, with covariance matrix $\sigma^2\bI$ . Notice that  $n_t$, the number of parameters involved in modeling $\bx(t)$, also known as order, can be variable in time and is seldom a priorly known. The aim is mainly to estimate $n_t$ and the position parameters ($\{\theta_k(t)\}$), since they often carry the desired information. However, since the model in \eqref{eq:model} is linear in the amplitude parameters, once the position parameters are replaced by their estimates, a standard linear estimator such as  Ordinary Least Squares (OLS) may be used to estimate the amplitudes. The problem of estimating $n_t$ is often called model order selection.

More formally, the observation model in \eqref{eq:model} can be written as
\begin{equation}\label{eq:obs_model1}
p(\bx(t)\mid S_t)=\frac{1}{(\pi\sigma^2)^m}e^{-\frac{\|\bx(t)-\suml_{k=1}^{n_t}\ba (\theta_k(t))s_k(t)
\|_2^2}{\sigma^2}}
\end{equation}
where the finite set $S_t$, given by
\begin{equation}\label{eq:param}
S_t=\{(\theta_1(t),s_1(t)),(\theta_2(t),s_2(t)),\ldots,(\theta_{n_t}(t),s_{n_t}(t))\}.
\end{equation}
represents the state. We further assume that the amplitudes $s_k(t)$ are distributed by a centered Gaussian pdf with variance $\calI_k(t)$, which we refer to as intensity. This can be formally written as
\begin{equation}\label{eq:hyerarchy}
p(s_k\mid\calI_k(t))=\frac{1}{\pi\calI_k(t)}e^{-\frac{|s_k|^2}{\calI_k(t)}}
\end{equation}
After straightforward manipulations, combining \eqref{eq:obs_model1}  and \eqref{eq:hyerarchy}, and integrating over $s_k$ leads to the following likelihood function in terms of the position and intensity parameters.
\begin{equation}\label{eq:obs_model_exact}
p(\bx(t)\mid \barS_t)=\frac{1}{\pi^m\det(\bR(t))}e^{-\bx^H(t)\bR^{-1}(t)\bx(t)}
\end{equation}
where
\begin{equation}
\barS_t=\{(\theta_1(t),\calI_1(t)),(\theta_2(t),\calI_2(t)),\ldots,(\theta_{n_t}(t),\calI_{n_t}(t))\}
\end{equation}
is a new state representation, here called the hyper-state, and
\begin{equation}
\bR(t)=\bR(\barS_t)=\sigma^2\bI+\suml_{k=1}^{n_t}\calI_k(t)\ba(\theta_k(t))\ba^H(\theta_k(t))
\end{equation}
The recent findings \cite{stoica2012spice} in the field of sparsity-based estimation suggest to substitute the determinant term with an exponential function to obtain 
\begin{equation}\label{eq:obs_model}
p(\bx(t)\mid \barS_t)\propto e^{-\bx^H(t)\bR^{-1}(t)\bx(t)-\lambda_0\suml_k\calI_k}
\end{equation}
where $\lambda_0$ is related to the average number of parameters, and practically treated as a design parameter.  Considered in this work, the model in \eqref{eq:obs_model} leads to a convex ML estimator, known as SParse Iterative Covariance-based Estimation (SPICE). Moreover, the convexity of the negative log-likelihood leads to unimodality of the posterior distributions.

\subsection{Time Evolution Model}
\label{sec:problem:evolution}

For the applications of interest herein, it is not suitable to consider an evolution model for the state $S_t$. Instead, a motion model for the hyperstate $\barS_t$ is considered. The motion model is a Markov chain, represented by the transitional probability density $Q (\barS,\barS^\prime)=p(\barS_{t+1}=\barS\mid \barS_t=\barS^\prime)$. It assigns to any pair of finite sets $(\barS,\barS^\prime)$ a value, quantifying the likelihood of $\barS^\prime$ being followed by $\barS$. Note that we consider a temporally constant transition function $Q$, corresponding to a stationary Markov Chain (MC). Then,
the joint p.d.f. of the sequence of state sets over an arbitrary window $\{t_1,t_1+1,\ldots,t_2\}$ of time is given by 
\begin{eqnarray}\label{eq:Markov}
&p(\barS_{t_1},\barS_{t_1+1},\barS_{t_1+2}\ldots,\barS_{t_2})=\nonumber\\
&p_{t_1}(\barS_{t_1})Q(\barS_{t_1+1},\barS_{t_1})
Q(\barS_{t_1+2},\barS_{t_1+1})\ldots Q(\barS_{t_2},\barS_{t_2-1})
\end{eqnarray} 
where $p_{t_1}(\barS_{t_1})$ denotes the marginal state distribution at the initial time $t_1$. We focus on a specific transition probability, associated with a case, where the elements of $S_{t}$ may first independently disappear with a small  probability $\alpha$. Then, the surviving elements may be modified by scalar models $p_0(\theta_{t+1}=\theta\mid\theta_t=\theta^\prime)$ , $p_1(\calI_{t+1}=\calI\mid\calI_t=\calI^\prime)$, and finally some new independent elements may be added according to a Poisson process with the hypothesis density function $\delta(\theta,\calI)$. This means that a new parameter may independently appear in a small neighborhood $\calN$ of a point $(\theta,\calI)$ with probability $\delta(\theta,\calI)\td(\calN)$, where $\td(\calN)$  is the volume (Lebesgue measure) of $\calN$. Note that 
\begin{equation}
\delta=\intl_{\Theta \times \bbR_+}\delta(\theta,\calI)\td\theta\td\calI<\infty.
\end{equation}
is the average rate of parameter birth, here assumed to be small. 
Then, the transition probability $Q(\barS,\barS^\prime)$ is given by
\begin{eqnarray}\label{eq:transition}
&Q(\barS_{t+1}=\barS,\barS_t=\barS^\prime)=e^{-\delta}\suml_{R\in \calT(\barS,\barS^\prime)}\nonumber\\
&\alpha^{|\barS^\prime|-|R|}(1-\alpha)^{|R|}\prodl_{(\theta,\calI,\theta^\prime,\calI^\prime)\in R}p_0(\theta\mid\theta^\prime)p_1(\calI\mid\calI^\prime)\prodl_{\theta\notin d_1(R)}\delta(\theta)
\end{eqnarray} 
where each summand is defined by an assignment $R$ between the elements of $\barS$ and the elements of $\barS^\prime$. Note that $|\barS^\prime|-|R|$ is the number of removed parameters from $\barS^\prime$, and the set ${\theta\notin d_1(R)}$ contains the newly introduced parameters in $\barS$. Hence, the three product terms in the summand evaluate the probabilities of survival, alteration and birth, respectively and according to the assignment $R$. The question of interest herein is to provide a filter, estimating the set $\barS_t$ at each time $t$ based on the observations $\bx(1),\bx(2),\ldots,\bx(t)$, the observation model in \eqref{eq:obs_model} and the MC motion model given by the transition probability in \eqref{eq:transition}.


\section{Recursive Bayesian Filtering}
\label{sec:RPS}
The model in \eqref{eq:Markov} enables us to solve exactly the desired estimation problem in a recursive way. Denoting $X^{(t)}=[\bx(1),\bx(2),\ldots,\bx(t)]$, we observe that the best estimate, in the Maximum A Posterior (MAP) density, for $\barS_t$ based on the observations up to time $t$, i.e. $X^{(t)}$ is given by maximizing the conditional likelihood $p(\barS_t\mid X^{(t)})$. The special form of the MC model in \eqref{eq:Markov} allows to recursively calculate $p(\barS_t\mid X^{(t)})$ by applying the Bayes rule:
\begin{eqnarray}\label{eq:update}
p(\barS_t\mid X^{(t)})=\frac{p(\barS_t,\bx(t)\mid X^{(t-1)})}{p(\bx(t)\mid X^{(t-1)})}=\nonumber\\
\frac{p(\bx(t)\mid \barS_t)p(\barS_t\mid X^{(t-1)})}{\intl_{\calS}p(\bx(t)\mid \barS_t=\barS)p(\barS_t=\barS\mid X^{(t-1)})\td\barS}
\end{eqnarray}
where $\calS$ denotes the entire space of the hyper-states, discussed in Appendix \ref{appendix:Calculus}, and 
\begin{equation}\label{eq:prediction}
p(\barS_t\mid X^{(t-1)})=\intl_{\calS}Q(\barS_t,\barS_{t-1}=\barS)p(\barS_{t-1}=\barS\mid X^{(t-1)})\td \barS
\end{equation}
The resulting recursion is simple: Given the conditional distribution $p(\barS_{t-1}\mid X^{(t-1)})$  at time instant $t-1$, calculate the prediction distribution $p(\barS_{t}\mid X^{(t-1)})$ by \eqref{eq:prediction}. Then, use \eqref{eq:update} to update the conditional distribution to $p(\barS_{t}\mid X^{(t)})$. As seen, the denominator in \eqref{eq:update} is independent of $\barS_t$. Thus, it can be replaced by any other scaling factor, without affecting the final result of MAP estimation, simplifying the calculations. This is called recursive Bayesian filtering.

The difficulty in the above method is to store the conditional distribution and calculate the integral in \eqref{eq:prediction}. Our method here is to consider the following family of approximate distributions, parametrized by an arbitrary symmetric positive semidefinite matrix $\hbR$ and an arbitrary positive weight function $\lambda:\Theta\to\bbR_+$ as follows 
\begin{equation}\label{eq:approxim}
p(\barS;\ \hbR , \lambda)=\exp{-\tr(\hbR\bR^{-1})-\suml_{(\theta,\calI)\in\barS}\lambda(\theta)\calI}
\end{equation}
where 
\begin{equation}
\bR=\bR(\barS)=\sigma^2\bI+\suml_{(\theta,\calI)\in\barS}\calI\ba(\theta)\ba^H(\theta)
\end{equation}
We approximate the posteriors by selecting the closest distribution in the KL sense in this family. We denote the parameters of the closest distribution to $p(\barS_t\mid X^{(t-1)})$ and $p(\barS_t\mid X^{(t)})$ by ($\hbR_t^-,\lambda_t^-$) and ($\hbR_t^+,\lambda_t^+$),  respectively.

The distribution in \eqref{eq:approxim} is necessarily unimodal. Moreover, when $\hbR$ and $\lambda$ are large, it is highly concentrated around its global maximal point, called the Maximum A Posterior (MAP) hyper-state estimate. When the updated distribution $p(\barS_t\mid X^{(t)})$ is considered, the resulting MAP estimate is the filter output (the desired estimate). When, $p(\barS_t\mid X^{(t-1)})$ is instead considered, the MAP estimate is called the predicted hyper state.

\subsection{Calculating the MAP Hyper-State Estimate}
One of the advantages with the above choice of approximate distribution is that it simplifies calculating the maximum a posterior estimate. When the posterior $p(\barS_t\mid X^{(t)})$ is calculated and approximated by parameters ($\hbR_t^+,\lambda_t^+$), the hyper-state MAP estimate is calculated by
\begin{equation}\label{eq:MAPestim}
\hat{\barS}_t=\arg\maxl_{\barS\in\calS}p(\barS_t\mid \hbR_t^+,\lambda_t^+)
\end{equation}
Similarly, the MAP predicted hyper-state is defined by
\begin{equation}\label{eq:MAPpredict}
\hat{\barS}^-_t=\arg\maxl_{\barS\in\calS}p(\barS_t\mid \hbR_t^-,\lambda_t^-)
\end{equation}
Both optimizations in \eqref{eq:MAPpredict} and \eqref{eq:MAPestim} yield to
\begin{eqnarray}\label{eq:SPICE}
&\hat{\barS}_t=\arg\minl_{\barS\in\calS}\nonumber\\
&\tr\left[\left(\sigma^2\bI+\suml_{(\theta,\calI)\in\barS}\calI\ba(\theta)\ba^H(\theta)\right)^{-1}\hbR^\pm_t\right]+
\suml_{(\theta,\calI)\in\barS}\calI\lambda_t^\pm(\theta)\nonumber\\
&\ 
\end{eqnarray}
where the plus and negative sign is for \eqref{eq:MAPestim} and  \eqref{eq:MAPpredict}, respectively. The optimization in \eqref{eq:SPICE} is a type of sparsity-based estimator and can be solved fast and precisely, with the so called weighted SPICE technique. First, a fine grid $\{\tiltheta_1,\tiltheta_2,\ldots,\tiltheta_N\}$ over  $\Theta$ is considered. Then, the following convex optimization is solved and the non-zero elements are selected as the estimates. 
\begin{eqnarray}\label{eq:SPICE_vec}
&\minl_{(\tlcalI_1,\tlcalI_2,\ldots,\tlcalI_N) \geq 0}\nonumber\\
&\tr\left[\left(\sigma^2\bI+\suml_{k=1}^N\tlcalI_k\ba(\tiltheta_k)\ba^H(\tiltheta_k)\right)^{-1}\hbR^\pm_t\right]+
\suml_{k=1}^N\tlcalI_k\lambda_t^\pm(\tiltheta_k)\nonumber\\
&\ 
\end{eqnarray}
The optimization in \eqref{eq:SPICE_vec} can be solved either by the off-the-shelf techniques, such as the CVX toolbox, or by the specific technique explained in \cite{stoica2011spice}. 
 
\subsection{Update Step}
Assume that at a certain time instant $t$, the posterior $p(\barS_t\mid X^{(t-1)})$ is approximated by parameters $(\hbR_t^-,\lambda_t^-)$. Once the vector $\bx(t)$ is observed, the posterior is changed according to  \eqref{eq:update}, which using \eqref{eq:obs_model}, results in 
\begin{eqnarray}\label{eq:lambda_upd1}
&p(\barS_t\mid X^{(t)})\propto\nonumber\\ &e^{-\bx^H(t)\bR^{-1}(t)\bx(t)-\tr(\hbR_t^-\bR^{-1}(t))-\suml_{(\theta,\calI)\in\barS_t}(\lambda_t^-(\theta)\calI+\lambda_0\calI)}=
\nonumber\\
&\exp\left\{-\tr\left[\bR^{-1}(t)\left(\hbR_t^-+\bx(t)\bx^H(t)\right)\right]\right.
\nonumber\\
&\left.-\suml_{(\theta,\calI)\in\barS_t}(\lambda_t^-(\theta)+\lambda_0)\calI\right\}
\end{eqnarray}
 We obtain that
\begin{eqnarray}\label{eq:update:final}
&\hbR^+_t=\hbR^-_t+\bx(t)\bx^H(t)\nonumber\\
&\lambda^+_t(\theta)=\lambda_t^-(\theta)+\lambda_0
\end{eqnarray}
\subsection{Prediction Step Approximation}
Now, consider occasions where the posterior $p(\barS_{t+1}\mid X^{(t)})$ is to be calculated by \eqref{eq:prediction}. Assume that the posterior $p(\barS_{t}\mid X^{(t)})$ is approximated by the parameters $\hbR_t^+$ and $\lambda^+_t$, and that these parameters are large enough, such that the corresponding posterior is highly concentrated around the filter output $\hat{\barS}_t$. In this case and according to Appendix \ref{appendix:local}, $\barS_{t}$ is a result of perturbing the  parameters of the MAP hyper-state estimate with a Gaussian perturbation, followed by adding  extra elements $(\theta^f_k,\calI^f_k)$, distributed by a Poisson  distribution. For simplicity, let us denote  $\hat{\barS}_t=\{(\theta_1,\calI_1),\ldots,(\theta_n,\calI_n)\}$ and denote by $\Delta\theta_k$ and $\Delta\calI_k$ the perturbations in $\theta_k$ and $\calI_k$, respectively. Then, according to the extended Laplace's method, derived in Appendix \ref{appendix:local:Laplace}, we may approximate $p(\barS_{t}\mid X^{(t)})$ :
\begin{eqnarray}\label{eq:laplace}
&\Delta\theta_k\sim\calN(\mathbf{0},G_k^{-1}),\ \Delta\calI_k\sim\calN(\mathbf{0},H_k^{-1})\nonumber\\
&\{(\theta^f_k,\calI^f_k)\}\sim\text{Poisson}(\omega(\theta,\calI))
\end{eqnarray}
where
\begin{eqnarray}
&G_k=\frac{\partial^2\tr(\hbR^+_t\bR^{-1})}{\partial\theta_k^2},\quad H_k=\frac{\partial^2\tr(\hbR^+_t\bR^{-1})}{\partial\calI_k^2}
\nonumber\\
&\omega(\theta,\calI)=\exp\left(\frac{\ba(\theta)^H\bR_+^{-1}(t)\hbR_t^+\bR_+^{-1}(t)\ba(\theta)\calI}
{1+\ba(\theta)^H\bR_+^{-1}(t)\ba(\theta)\calI}-\lambda^+_t(\theta)\calI\right)
\end{eqnarray}
Simple  calculations show that after applying time evolution by \eqref{eq:prediction}, the approximation in \eqref{eq:laplace} still holds, but the parameters $G_k,H_k$ and $\omega(\theta,\calI)$ are updated (See \cite{mahler2000theoretical} for the Poisson Process under time evolution) to 
\begin{eqnarray}
&G^\prime_k=\frac{G_k}{1+\sigma_\theta^2G_k},\quad H^\prime_k=\frac{H_k}{1+\sigma_{\calI}^2H_k}
\nonumber\\
&\omega^\prime(\theta,\calI)=(1-\alpha)\intl_{\Theta\times\bbR_+}\omega(\theta^\prime,\calI^\prime)p_0(\theta\mid\theta^\prime)
p_1(\calI\mid\calI^\prime)\td\theta\td\calI+\nonumber\\
&\delta(\theta,\calI)
\end{eqnarray}
respectively, where $\sigma_\theta^2$ and $\sigma_\calI^2$ are the perturbation variance, associated with the time evolution models $p_0$ and $p_1$, given by $\var(\theta_t\mid\theta_t=\theta_k)$ and $\var(\calI_t\mid\calI_t=\calI_k)$ respectively. This represents the posterior distribution after time evolution. Now, we project this distribution on the desired space of parametrized distributions by $\hbR$ and $\lambda$. We perform this by taking the minimum KL distance. Although the process is generally intractable, assuming that time evolution is small, i.e. the hyper-state does not change fast, the process can be easily performed by perturbation theory. Appendix \ref{appendix:perturb}, establishes this relation. Here, we consider the final result, where limited computational complexity is also considered. The simplified prediction steps can be represented by
\begin{eqnarray}\label{eq:prediction:final}
&\hbR^-_{t+1}=\frac{\suml_k\frac{1}{1+\sigma_\theta^2G_k}+\suml_k\frac{1}{1+\sigma_\calI^2H_k}}{2n}\hbR^+_{t}\nonumber\\
&\lambda^-_{t+1}(\theta)=\left[\lambda^+_{t}(\theta)-\right.\nonumber\\
&\left.\frac{\delta_1(\theta)}{2}(\lambda(\theta)-\ba^H(\theta)\bR_+^{-1}(t)\hbR^+_t\bR_+^{-1}(t)\ba(\theta))\right]_+
\end{eqnarray}   
where   $\delta_1(\theta)=\intl_{\bbR_+}\calI\delta(\theta,\calI)\td\calI$. The overall proposed algorithm is summarized in Algorithm \ref{alg1}.
\begin{algorithm}
\begin{algorithmic}
\Require A positive definite matrix $\bC$ and a positive function $\delta_1(\theta)$.
\State Initialize by a symmetric positive definite matrix $\hbR_1^-$ and a positive function $\lambda_1^-(\theta)$.
\State Set $t=1$.
\Repeat

\State Observe $\bx(t)$ and calculate $\hbR_t^+$ and $\lambda_t^+$ from \eqref{eq:update:final}.

\State Calculate $\hat{\calS_t}$ by solving its corresponding SPICE optimization in \eqref{eq:SPICE_vec} and selecting nonzero elements. Calculate $\bR_+(t)=\bR(\hat{\calS_t})$.

\State Calculate $\hbR_{t+1}^-$ and $\lambda_{t+1}^-$ from \eqref{eq:prediction:final}.
 
\State Set $t\gets t+1$.

\Until{Required.}
\end{algorithmic}
\caption{The proposed algorithm.}
\label{alg1}
\end{algorithm}

\section{Numerical Results and Comparison to Related Works}\label{sec:anal}
In this section, we examine the method developed in Section \ref{sec:RPS} in a number of selected scenarios and compare the results on the synthetic data to other filtering technique. We consider the problem of Direction of Arrival (DOA) estimation with a Uniform Linear Array (ULA), where the position parameter is the direction (angle) of an electromagnetic source and the amplitude is the complex envelope of the electromagnetic wave transmitted by it and the observation vector is the signal measured at a ULA of $m=20$ sensors. The DOA is often reparametrized for simplicity, introducing the electrical angle, which we utilize here. Then, \eqref{eq:model} holds with
\begin{equation}
\ba(\theta)=[1\ e^{j\theta}\ e^{2j \theta}\ \ldots \ e^{(m-1)j\theta}]
\end{equation}
where $\theta\in\Theta=[-\pi\ \pi]$ is the electrical angle.

In all simulations, we use a Gaussian MC model for parameter evolution, i.e.
\begin{equation}\label{eq:GaussianMC}
p_{0}(\theta\mid\theta^\prime)=
\frac{1}{\sqrt{2\pi\sigma_\theta^2}}e^{-\frac{(\theta-\theta^\prime)^2}{2\sigma_\theta^2}}
\end{equation}
and
\begin{equation}\label{eq:GaussianMC2}
p_{1}(\calI\mid\calI^\prime)=
\frac{1}{\sqrt{2\pi\sigma_\calI^2}}e^{-\frac{(\calI-\calI^\prime)^2}{2\sigma_\calI^2}}
\end{equation}

We also perform the calculations over the spectra (e.g. $\lambda(\theta)$) in the recursive algorithms of interest, by taking a uniform grid $\tilTheta$ over $\Theta$ with minimum separation $0.01$. This results in $629$ grid points.  The average false alarm power $\delta_1(\theta)$ is also selected uniformly over $\Theta$, i.e. $\delta(\theta)=\delta$. 

\subsection{Related Studies}\label{sec:prior1}

In the literature, there is a number of different studied approaches, applicable to the problem of interest herein. We briefly review some of the more popular ones, considered her for comparison.

\subsubsection{Sliding Window Techniques}
\label{sec:prior1:Sliding}
In the simplest case, a temporal window is considered, which is generally defined by a window function $w_{\tau}$ for $\tau= 0,1,\ldots$. At a given time $t$, the following optimization is solved  
\begin{eqnarray}\label{eq:MLbatch}
&(\htheta_1(t),\htheta_2(t),\ldots,\htheta_n(t))=
\arg\minl_{\theta_1,\theta_2,\ldots,\theta_n}
\minl_{s_1(t),s_2(t),\ldots,s_n(t)}\nonumber\\
&\suml_{\tau=1}^T
w_{\tau}\left\|\bx(t-\tau)-\suml_{k=1}^n\ba(\theta_k)s_k(t-\tau)\right\|_2^2\nonumber\\
&\ 
\end{eqnarray}
Then, the position parameters $\htheta_k(t)$ of the global minimum point is the filter output. Notice that the summation in the cost of \eqref{eq:MLbatch} is over time, but the parameters $\theta_k$ are not time dependent. The motivation for \eqref{eq:MLbatch} is that the error in assuming constant position parameters can be approximately modeled by the increase in the noise variance with the factors $\{w_{\Delta t}\}$. Optimizing \eqref{eq:MLbatch} is equivalent to solving the ML estimator for such an approximate model. Also, note that the  order $n$ is fixed. In practice, where the order is typically unknown and variable, \eqref{eq:MLbatch} is solved for a variety of orders. This can be efficiently done, e.g. by the RELAX technique \cite{li1996efficient}. Denoting by $V_n$ the optimal value of \eqref{eq:MLbatch}, the order and its corresponding solution is selected by a rule over the collection $\{V_n\}$, generally called information criterion. We consider a popular choice of information criterion, given by minimizing 
\begin{equation}\label{eq:IC}
\minl_n V_n+kn
\end{equation}
where $k$ is a design parameter. The choice of $k$ for asymptotic cases and other information criteria are discussed in \cite{GLRT_IC}. When $w_{\Delta t} =\delta_{0,\Delta t}$, i.e it is non zero, only when $\Delta=0$, the optimization in \eqref{eq:MLbatch} simplifies to the exact ML estimator based on the observation model. We refer to this as the "instantaneous" estimator. 

The inner optimization in \eqref{eq:MLbatch} can be solved analytically to obtain 
\begin{equation}\label{eq:MLbatch2}
(\htheta_1,\htheta_2,\ldots,\htheta_n)=\arg\max_{\theta_1,\theta_2,\ldots,\theta_n}
\tr\left(\hbR_t\bP_{\bA(\theta_1,\ldots,\theta_n)}\right)
\end{equation}
where $\hbR_t=\suml_{\Delta t=1}^tw_{\Delta t}\bx(t-\Delta t)\bx^H(t-\Delta t)$ is the windowed sample correlation matrix, $\bA(\theta_1,\ldots,\theta_n)=[\ba(\theta_1),\ldots,\ba(\theta_n)]=\bA$ and $\bP_{\bA(\theta_1,\ldots,\theta_n)}=\bA(\bA^H\bA)^{-1}\bA^H$ is the projection matrix into the range space of $\bA$, also known as the signal space. Solving \eqref{eq:MLbatch2} is still difficult, but the following approximate technique can be used: First, the closest projection matrix $\hbP_T$ to $\hbR_T$ in the Frobenius distance is found as
\begin{equation}\label{eq:subspace_model}
\hbP_T=\bU_{n,T}\bU_{n,T}^H
\end{equation}
where $\bU_{n,T}$ is the collection of the eigenvectors related to the $n$ largest eigenvalues of $\hbR_T$. Then, the closest bases $\ba(\theta)$ to the range space of $\hbP_T$ is selected by taking the local minima of the spectrum $u_T(\theta)=\|\ba(\theta)-\hbP_T\ba(\theta)\|_2^2$. This technique is called MUltiple SIgnal Classification (MUSIC).

\subsubsection{Target Tracking Techniques}

From one perspective, the target tracking techniques are to enhance the quality of estimates provided by other methods, such as the instantaneous estimates. Suppose that an instantaneous estimator is utilized to obtain a preliminary set of estimates $Z_t=\{\htheta_1(t),\htheta_2(t),\ldots,\htheta_{\hn_t}(t)\}$. Then, the estimates can be related to $X_t$ through the analysis of the instantaneous estimator, leading to a conditional pdf $p(Z_t\mid X_t)$. As seen, the resulting model is again RFS based. Most often, the following approximate relation, very similar to the evolution model in \eqref{eq:transition} is considered.
\begin{eqnarray}\label{eq:TargetDetection}
&p(Z_t\mid X_t)=e^{-\mu}\suml_{R\in \calT(Z_{t},X_t)}\nonumber\\
&\beta^{|R|}(1-\beta)^{|X_t|-|R|}\prodl_{(\htheta,\theta)\in R}p_1(\htheta\mid\theta)\prodl_{\theta\notin d_1(R)}\mu(\theta)
\end{eqnarray}
where $\beta$ is the probability of detection of a parameter, $p_1(\htheta\mid\theta)$ is the distribution of an estimates $\htheta$, corresponding to the true parameter $\theta$, and $\mu(\theta)$ is the hypothesis density for the false alarm (false detection) process, assumed to be a Poisson process. Note that
\begin{equation}
\mu=\intl_\Theta\mu(\theta)\td\theta<\infty
\end{equation}
is the average false alarm rate. Given \eqref{eq:transition} and \eqref{eq:TargetDetection}, we may use \eqref{eq:update} and \eqref{eq:prediction} to obtain a recursive filter, called target tracking filter. The exact result is generally numerically intractable. To maintain a limited amount of calculations in the course of target tracking, the method of Probability Hypothesis Density (PHD) \cite{mahler2000theoretical} approximates the resulting posterior distributions by the Poisson process, leading to the following steps: Denoting by $D^+_t$ and $D^-_t$, the PHDs for the updated and predicted posteriors, respectively, the prediction in \eqref{eq:prediction} is exactly resolved to give
\begin{equation}\label{eq:PHDpredict}
D^+_t(\theta)=\alpha\intl_\Theta p_0(\theta\mid\theta^\prime)D^-_{t-1}(\theta^\prime)\td\theta^\prime+\delta(\theta)
\end{equation}
and the closest approximation in the Kullback-Leibler sense to the result of the calculations in \eqref{eq:update} is found to be
\begin{equation}\label{eq:PHDupdate}
D^-_t(\theta)=(1-\beta)D^+_t(\theta)+\suml_{\htheta\in Z_t}\frac{\beta p_1(\htheta\mid\theta)D^+_t(\theta)}{\beta \intl_\Theta p_1(\htheta\mid\theta)D^+_t(\theta)\td\theta+\mu(\htheta)}
\end{equation}
The final estimates are given by local maxima of $D_t^-(\theta)$.

\subsubsection{Subspace-Based Techniques}
Another type of recursive filters is introduced, based on the subspace techniques such as the previously introduced MUSIC method. The idea is to replace $X_t=\{\theta_k(t)\}$ by the subspace $\mathscr{X_t}$, spanned by the bases $\{\ba(\theta_k(t))\}$. The subspace is represented by a projection matrix $\bP(t)$. An effective way to estimate $\bP(t)$, also considered here is to solve
\begin{equation}\label{eq:subspace}
\bP(t)=\arg\minl_\bP\|\bx(t)-\bP\bx(t)\|_2^2+\alpha\|\bP-\bP(t-1)\|_F^2
\end{equation}  
where $\bP(t-1)$ is the estimate at the previous time instant and $\alpha$ is a design parameter. Once $\bP(t)$ is calculated, the parameter estimates are obtained by the MUSIC technique. Note that this technique is loosely tied to the statistical model, stated in Section \ref{sec:problem}, though it enjoys a remarkably low computational complexity.

\subsection{ Numerical Results}
Now, we consider the introduced techniques and the proposed one in some scenarios. For the PHD observation model, we also choose
\begin{equation}
p(\htheta\mid\theta)=\frac{1}{\sqrt{2\pi\sigma_e^2}}e^{-\frac{(\htheta-\theta)^2}{2\sigma_e^2}}
\end{equation}   
where we treat $\sigma_e$ as a tuning parameter. The instantaneous estimator for the target tracking technique is RELAX with the information criterion in \eqref{eq:IC} and $k=3$. 

\subsubsection{Two Crossing Targets}
In this setup, two moving sources $(\theta_1,(t),\theta_2(t))$ were considered. They moved according to the equations $\theta_1=-\pi/2+0.01\pi t$ and $\theta_2=\pi/2-0.01\pi t$ for $t=1,2,\ldots, T=100$. Their corresponding amplitudes were randomly generated  by the standard Gaussian distribution. The noisy observations were obtained by adding centered, uncorrelated Gaussian noise to the observations, with variance $0.25$, providing SNR$\approx 6$dB.

The proposed technique was applied by $\lambda=2$ and $\sigma=0.5$, together with the time evolution parameters $\delta_1(\theta)=0.1$ and $\sigma_\theta=\sigma_\calI=0.03$. We also considered instantaneous estimation by RELAX and enhanced the results by PHD filtering. For the latter case, the parameters $\beta=0.99$, $\alpha=0.01$, $\delta(\theta)=10^{-4}$, $\mu=0.04$ and $\sigma_e=0.01$ are selected. Moreover, the subspace technique in \eqref{eq:subspace} is used with $\alpha=2$, adjusted for the best result.

\begin{figure}[!t]
\centering
\includegraphics[height=5cm, width=9cm]{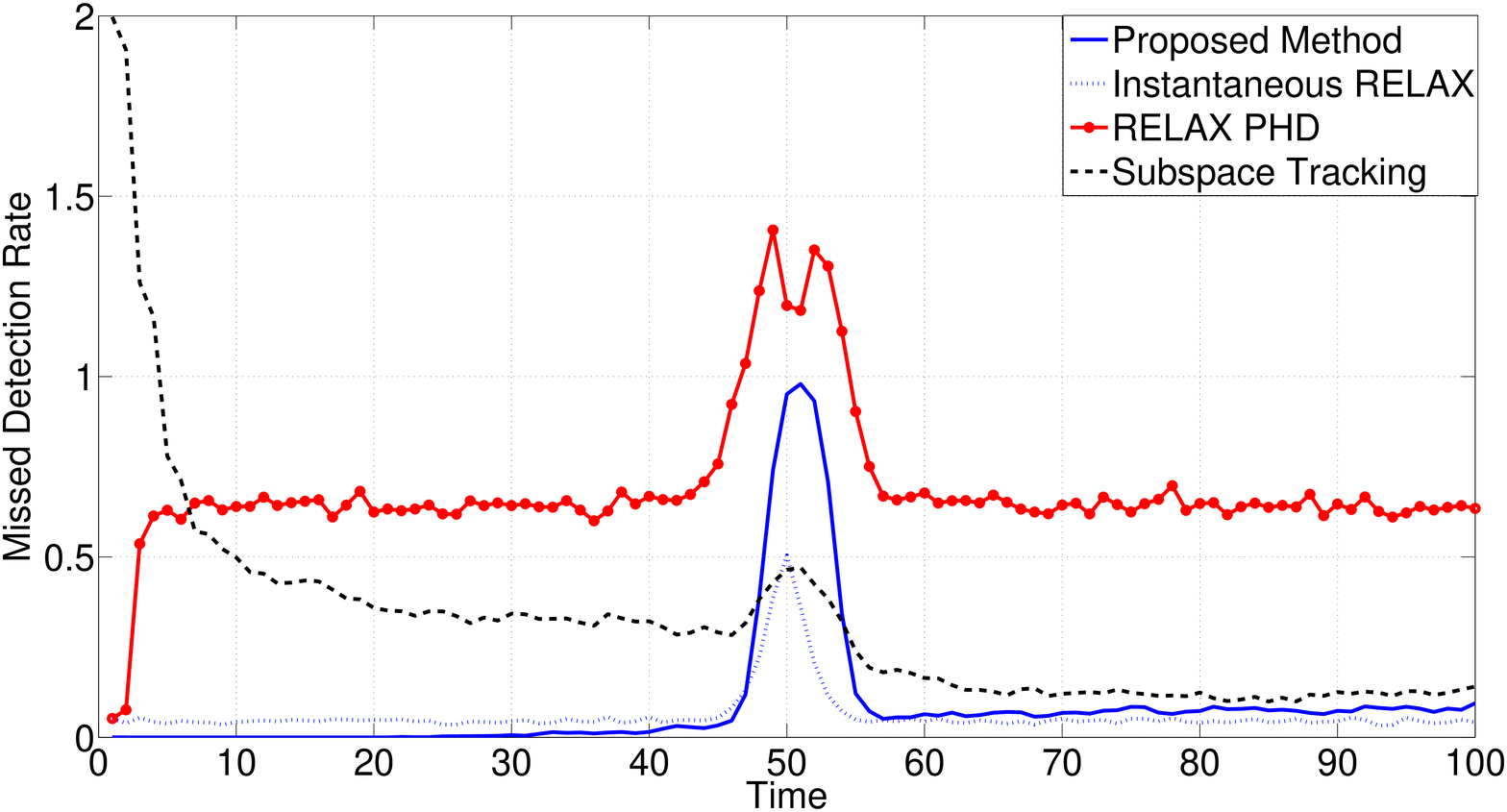}
\caption{Missed detection rate in the deterministic crossing setup, averaged over 16000 trials. }
\label{fig:MT_MD_LQ}
\end{figure}

\begin{figure}[!t]
\centering
\includegraphics[height=5cm, width=9cm]{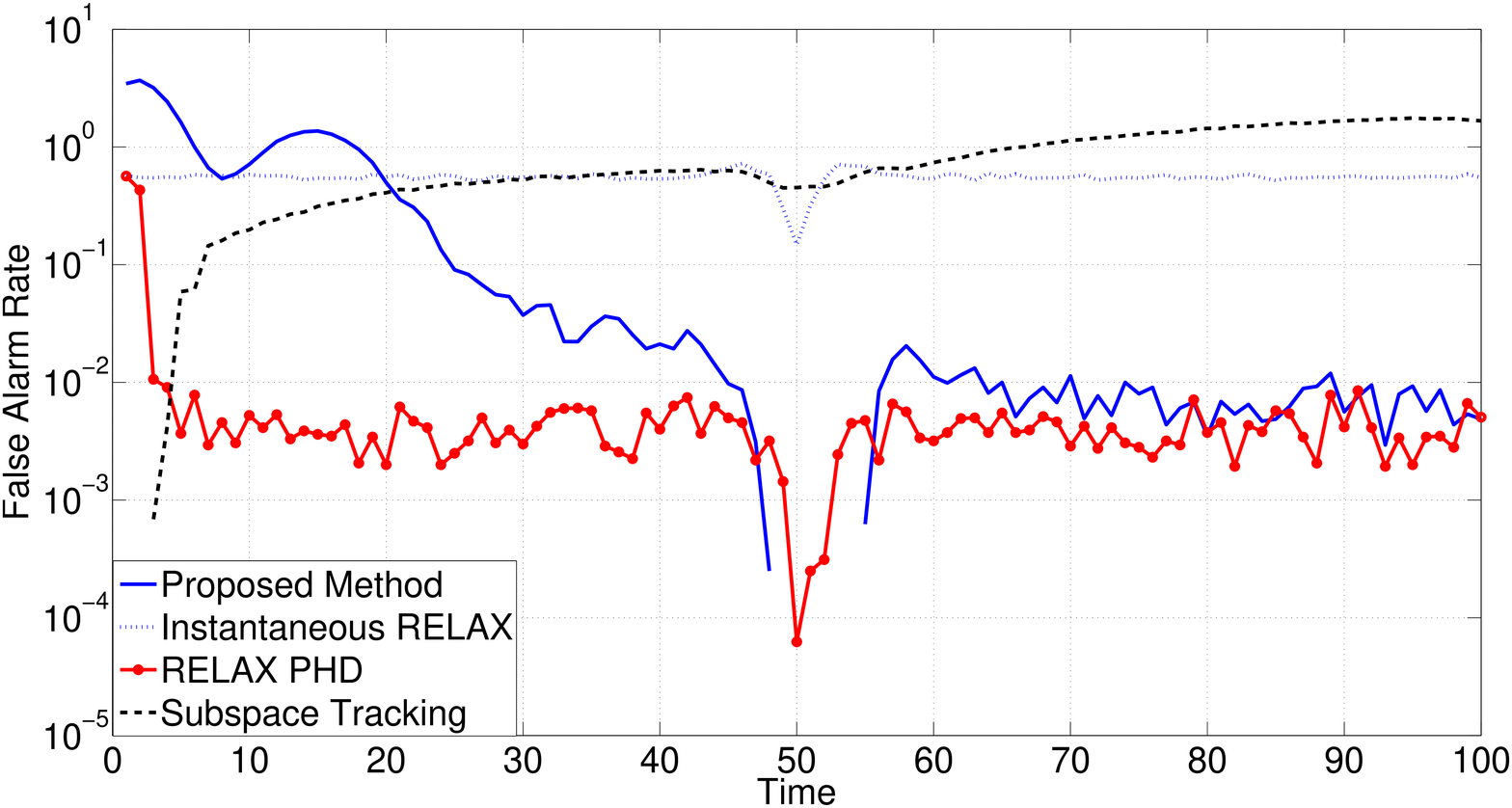}
\caption{False alarm rate in the deterministic crossing setup, averaged over 16000 trials.}
\label{fig:MT_FA_LQ}
\end{figure}

\begin{figure}[!t]
\centering
\includegraphics[height=5cm, width=9cm]{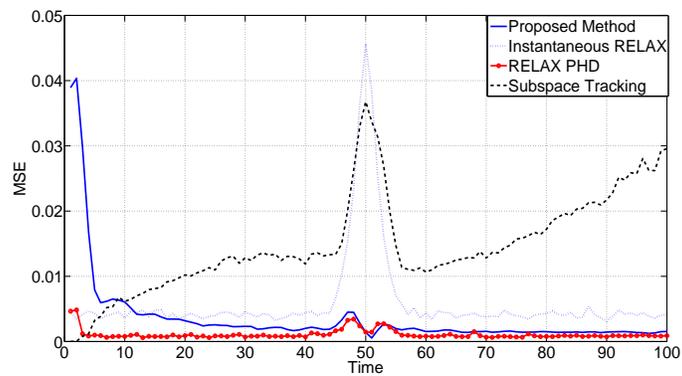}
\caption{Mean square error in the deterministic crossing setup, averaged over 16000 trials.}
\label{fig:MT_ER_LQ}
\end{figure}

In terms of missed detection, false alarm and error, figures \ref{fig:MT_MD_LQ}, \ref{fig:MT_FA_LQ} and \ref{fig:MT_ER_LQ} depict the average quality of the resulting estimates over time, respectively. At a specific time, the number of false alarms, and missed detections are simply calculated as the number of exceeding or lacking parameters, namely $(\hat{n}_t-n_t)_+$ and $(n_t-\hat{n}_t)_+$, respectively. The error is calculated by adding the square error over the best assignment between estimates and the true parameters.

As seen, the instantaneous RELAX estimator typically has a high false alarm rate. The PHD filter substantially improve both the false alarm, and the error properties of the RELAX method, but increases the missed detection rate. Changing the parameters of the PHD filter modifies the trade off between false alarm and missed detection, but may not improve both. On the other hand, the proposed technique has improved miss-detection properties, but slightly increases the error level. This is due to the mismatch between the exact model in \eqref{eq:model} and the applied one in \eqref{eq:obs_model}, which is well known to result in biased estimates. It is clearly seen that the proposed technique initially needs about 40 samples to achieve its steady behavior, but later rapidly adapts itself to a varying environment. This may imply an improper choice of initial parameters. Finally, notice that the proposed technique provides better results at the crossing point, suggesting that the proposed method relies more on the time correlation of parameters. This can also be seen from the fact that in Figure \ref{fig:MT_ER_LQ}, the proposed technique corresponds to a smoother curve than the other techniques, showing higher temporal correlation between the estimates.     

 
\subsubsection{Single Target with a Sudden Change}

In a different setup, we considered a single target $\theta$. The target is assumed to be at rest for the first 100 samples, i.e. $\theta(t)=-\pi/2$ for $t=1,\ldots,100$. Next, it started a linear movement with an impulsive initial position change, given by $\theta(t)=3*\pi/2-0.01*\pi*t$ for $t=101,\ldots,2000$.

The proposed technique was compared to sliding window, with the window function $w_{\tau}=\eta^{\tau}$. This choice generally simplifies the calculations, since it leads to a recursive evaluation of the matrix $\hbR_t$ as
\begin{equation}\label{eq:results:mid1}
\hbR_{t+1}=\eta\hbR_t+\bx(t)\bx(t)^T
\end{equation}
where $\hbR_t$ is defined in \eqref{eq:MLbatch2}. It is interesting to see that the overall recursive calculation of $\hbR^+_t$ in the proposed algorithm is similar to \eqref{eq:results:mid1}, when the forgetting factor is replaced by a time-varying parameter. We also used the SPICE technique to solve \eqref{eq:MLbatch2} or equivalently \eqref{eq:MLbatch}, leading to the same optimization in \eqref{eq:SPICE_vec}, when $\hbR^\pm_t$ and $\lambda(\theta)$ are replaced by $\hbR_t$ and $\lambda_0/(1-\eta)$, respectively. From this perspective, the proposed method is a sliding window technique with a SPICE estimator, where adaptive forgetting factor and weights are utilized.

\begin{figure}[!t]
\centering
\includegraphics[height=5cm, width=9cm]{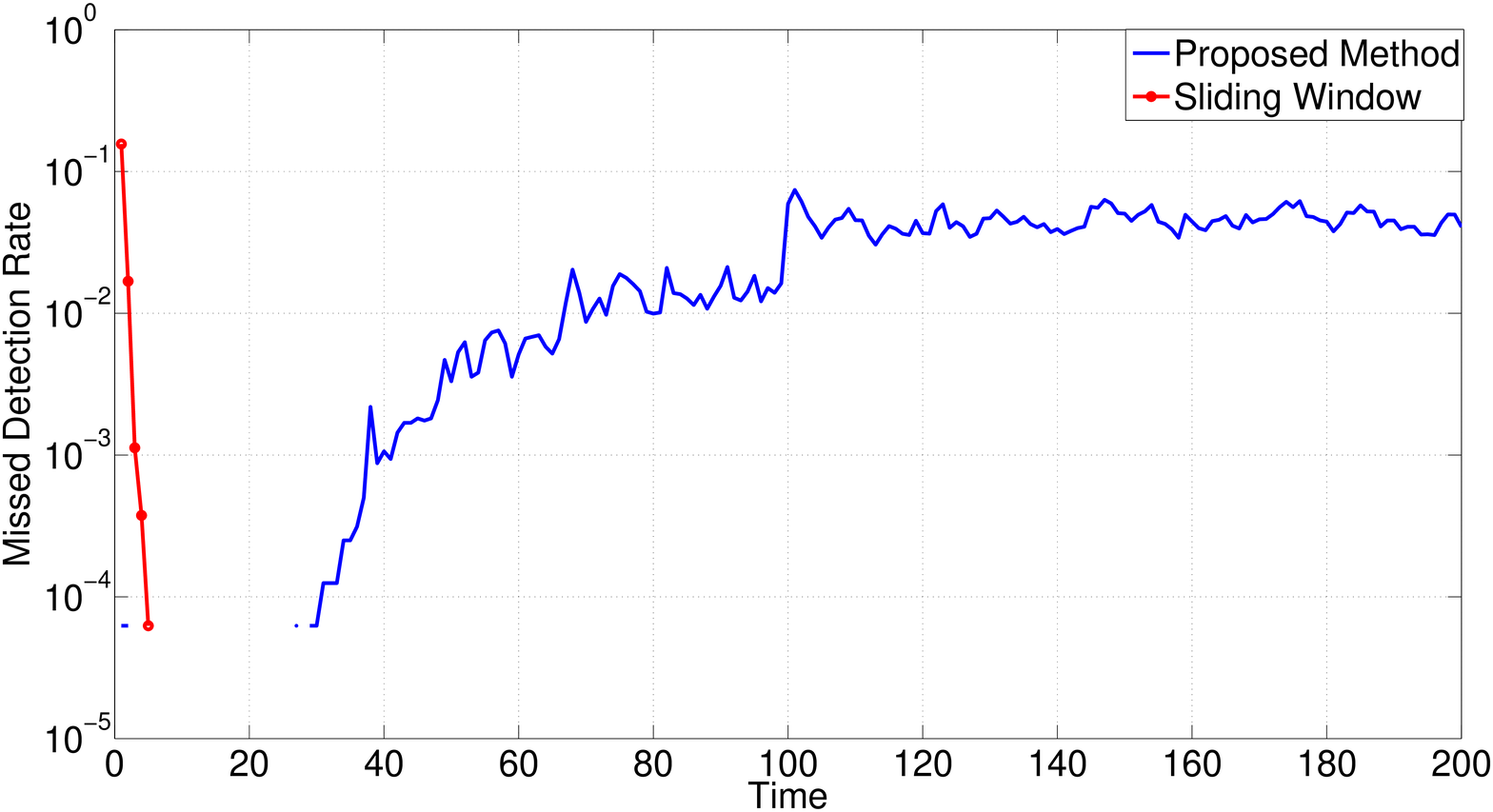}
\caption{Missed Detection rate in the sudden movement setup, averaged over 16000 trials.}
\label{fig:JT_MD_LQ}
\end{figure}

\begin{figure}[!t]
\centering
\includegraphics[height=5cm, width=9cm]{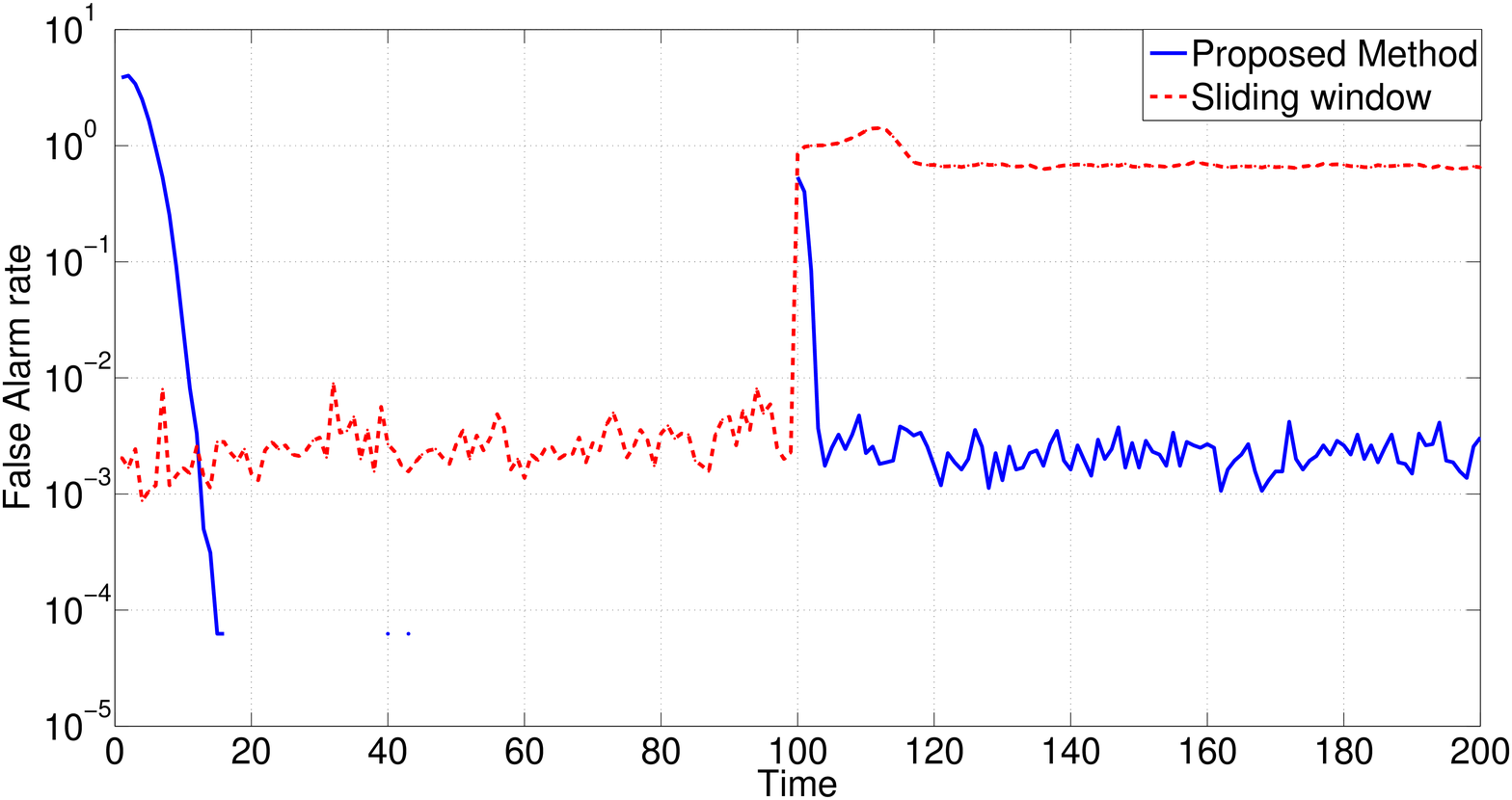}
\caption{False alarm rate in the sudden movement setup, averaged over 16000 trials.}
\label{fig:JT_FA_LQ}
\end{figure}

\begin{figure}[!t]
\centering
\includegraphics[height=5cm, width=9cm]{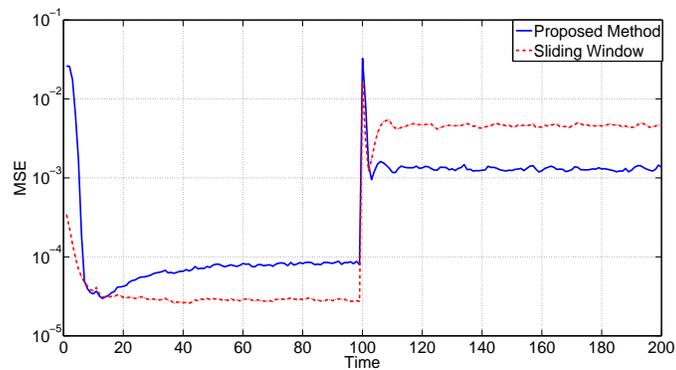}
\caption{Mean square error in the in the sudden movement setup, averaged over 16000 trials.}
\label{fig:JT_ER_LQ}
\end{figure}

Figures \ref{fig:JT_MD_LQ}, \ref{fig:JT_FA_LQ}, \ref{fig:JT_ER_LQ} depict the average missed detection, false alarm and error results, respectively, where the same parameters as the previous setup and $\eta=0.8$ were used. In the initial stationary phase, the sliding window technique outperforms the proposed one, since the setup fits the assumptions of the sliding window.   In terms of error, both techniques rapidly adapt to the sudden change, but the sliding window has a longer transient in terms of false alarm rate. 

\section{Concluding Remarks}
In this paper, the problem of filtering a variable number of parameters in difficult scenarios was discussed. We used a recent modified Bayesian model in \cite{stoica2012spice} and related it to a RFS-based evolution model to obtain a consistent representation for our problem of interest. Next, we approximated the corresponding recursive Bayesian filter to our model, and obtained a tractable filter. We simplified the design to avoid heavy computations. This led to a filter based on two components; An approximate data covariance matrix, and a weight function, controlling miss-detection over the space of parameters.

As the numerical experiments suggest, the technique is more robust to observation impairments and is more flexible against rapid movements. Our approach exploits, and is highly connected to the SPICE technique. Hence, it exhibits similar behavior. For example, it has a relatively short convergence rate and provides consistent estimates, but the effect of noise is not symmetric on the estimates. Mathematically speaking, the estimates have a small statistical bias, proportional to the noise power. The method also exhibits a robust behavior in a low SNR regime.  

Herein, the emphasis was on simplifying calculations at each recursion by avoiding difficulties with the grid-based spectral manipulations and instead combining approximate information of different time instants to maintain performance. As observed by simulations, this is favorable in a low SNR case, where fusing multiple observations is necessary to obtain a reliable estimate. However, the method might be improved if complexity is not an immediate concern and a more complex approximation is desirable.  Moreover, the possibility of bootstrapping and the application of particle filters should not be ruled out.

\appendices
\section{Calculus of Random Finite Sets}
\label{appendix:Calculus}

\subsection{Functional Representation}
To perform RBF, we need to calculate posteriors over finite sets, involving integration over RFS densities. Here, we review how this can be accomplished. In general, the probability distributions over the set of all finite sets can be represented by a sequence of real functions. For example, the marginal state distribution $p_{t}(\barS_{t})$ may be represented by the function sequence  $\{q_t^{(n)}:\bbR^n\times\bbR_+^n\to\bbR_+\}$ defined by
\begin{equation}
q_t^{(n)}(\theta_1,\ldots,\theta_n,\calI_1,\ldots,\calI_n)=p_{t_1}(\barS_t=\{(\theta_1,\calI_1),\ldots,(\theta_n,\calI_n)\})
\end{equation} 
Note that the functions $q_t^{(n)}$ are symmetric under the permutation of the pairs $(\theta_k,\calI_k)$, since the state set is invariant under such a transform. Moreover, for a fixed $n$,
\begin{equation}
\intl_{\bbR^n\times\bbR_+^n}q_t^{(n)}(\theta_1,\ldots,\theta_n,\calI_1,\ldots,\calI_n)
\td^n\theta\td^n\calI=n!\times p(n_t=n)
\end{equation}
The reason is that the left hand side integration hits each set $S_t$ of order $n$ exactly $n!$ times by different permutations of parameters, but does not hit a set $S_t$ of a different order. In the same manner, the transition probability $Q$ can be expressed by the following sequence of functions
\begin{eqnarray}\label{eq:func_q}
&q^{(n,n^\prime)}
(\theta_1,\ldots,\theta_n,\calI_1,\ldots,\calI_n,\theta^\prime_1,\ldots,\theta^\prime_{n^\prime},
\calI^\prime_1,\ldots,\calI^\prime_{n^\prime})=\nonumber\\
&Q(\barS=\{(\theta_k,\calI_k)\},\barS^\prime=\{(\theta_k^\prime,\calI_k^\prime)\})
\end{eqnarray}  
\subsection{Integration}

In general integration over the set of random finite sets can be explained in terms of the above functional representation. Consider the marginal distribution over the step of finite sets $\barS_t$, represented by sequence of functions $q_t^{(n)}$ and take a function $f(\barS):\calS\to\bbR$. Then, we have that
\begin{equation}
\intl_{\calS}f(\barS)\td\barS=\suml_{n=0}^\infty\frac{1}{n!}\intl_{\Theta^n\times\bbR_+^n}
q^n(\theta_1,\ldots,\theta_n,\calI_1,\ldots,\calI_n)\td^n\theta\td^n\calI
\end{equation}
Notice how division by $n!$ cancels the aforementioned effect of multiple recalculation. Other integrations such as marginalization in \eqref{eq:prediction} can be carried out in a similar manner. For example, suppose that the posterior $p(\barS_t\mid X^{(t)})$ is represented by functions $q_0^{(n)}$ at a certain time $t$. Then, the integration in \eqref{eq:prediction} yields to
\begin{eqnarray}
&p(\barS_{t+1}=\{(\theta_k,\calI_k)\}\mid X^{(t)})=\nonumber\\
&\suml_{n^\prime=0}^\infty\frac{1}{n^\prime!}\intl_{\Theta^{n^\prime}\times\bbR_+^{n^\prime}}
q^{(n,n^\prime)}q_0^{n^\prime}(\theta^\prime_1,\ldots,\theta^\prime_{n^\prime},\calI^\prime_1,\ldots,\calI^\prime_{n^\prime})
\td^{n^\prime}\theta^\prime\td^{n^\prime}\calI^\prime
\end{eqnarray}
where the similar argument of $q^{(n,n^\prime)}$ to \eqref{eq:func_q} is neglected.

\section{RFS Local Approximation}
\label{appendix:local}
Consider a distribution in the family, given by \eqref{eq:approxim}, and suppose that the parameters $\hbR$ and $\lambda$ are large. Take $\hat{\barS}=\{(\theta_1,\calI_1),\ldots,(\theta_n,\calI_n)\}$ as the maximum probability point. Then, a large deviation from $\hat{\barS}$ leads to a considerable  probability reduction. Thus, we may assume that the deviation is small. Hence, local Taylor expansion can be applied. Note that a small deviation from the set $\barS$ includes small perturbations leading to a typical hyper-state set
\begin{equation}
\barS=\{(\theta_k+\Delta\theta_k,\calI_k+\Delta\calI_k)\}\cup
\{(\theta^f_1,\calI^f_1),\ldots(\theta^f_{n_f},\calI^f_{n_f})\}
\end{equation}
where the parameters, indexed with $f$ are additional. Furthermore, the parameters $\Delta\theta_k$, $\Delta\calI_k$ and $\calI^f_k$ are assumed to be small. The negative log-density function is written as
\begin{eqnarray}
&-\log p(\barS;\ \hbR,\lambda)=\nonumber\\
&\tr\left[\hbR\left(\sigma^2\bI+\suml_k(\calI_k+\Delta\calI_k)\ba(\theta_k+\Delta\theta_k)\ba^H(\theta_k+\Delta\theta_k)+\right.\right.
\nonumber\\
&\left.\left.\suml_k\calI^f_k\ba(\theta^f_k)\ba^H(\theta^f_k)\right)^{-1}\right]+
\nonumber\\
&\suml_k\lambda(\theta_k+\Delta\theta_k)(\calI_k+\Delta\calI_k)+\suml_k\lambda(\theta_k^f)\calI^f_k
\end{eqnarray} 
We may now apply the Taylor expansion.

\subsection{Poisson Approximation}

Due to the local minimality of $\hat{\barS}$, it turns out that the effect of $\Delta\theta_k$ and $\Delta\calI_k$ vanish up to the first order. This means the negative log-distribution can be written as
\begin{eqnarray}
&-\log p(\barS;\ \hbR,\lambda)=\nonumber\\
&\tr\left[\hbR\left(\bR_0+\suml_k\calI^f_k\ba(\theta^f_k)\ba^H(\theta^f_k)\right)^{-1}\right]+
\nonumber\\
&\suml_k\lambda(\theta_k)(\calI_k)+\suml_k\lambda(\theta_k^f)\calI^f_k
\end{eqnarray} 
where $\bR_0=\bR(\hat{\barS})$. Using the matrix inversion lemma and neglecting the cross-product terms $\calI^f_k\calI^f_l$, we obtain
\begin{eqnarray}
&-\log p(\barS;\ \hbR,\lambda)=\nonumber\\
&-\log p(\hat{\barS};\ \hbR,\lambda)+\suml_k\left(\lambda(\theta_k^f)\calI^f_k-\frac{\ba^H(\theta_k^f)\bR_0^{-1}\hbR\bR_0^{-1}\ba(\theta_k^f)\calI^f_k}
{1+\ba^H(\theta_k^f)\hbR\bR_0^{-1}\ba(\theta_k^f)\calI^f_k}\right)
\end{eqnarray} 
This shows that up to the first order, the behavior of the RFS can be identified by the Poisson process of additional elements $(\theta_k^f,\calI^f_k)$ with density
\begin{equation}
w(\theta,\calI)=e^{-\left(\lambda(\theta)\calI-\frac{\ba^H(\theta)\bR_0^{-1}\hbR\bR_0^{-1}\ba(\theta)\calI}
{1+\ba^H(\theta)\hbR\bR_0^{-1}\ba(\theta)\calI}\right)}
\end{equation}  
\subsection{Extended Laplace's Method}
\label{appendix:local:Laplace}
To capture the behavior of $\Delta\theta_k$ and $\Delta\calI_k$, we need to consider the higher order terms. However, we neglect the cross-product terms in favor of numerical simplicity, and according to the fact they are often smaller due to low coherency in the basis manifold. Then after straightforward calculations, we obtain that
\begin{eqnarray}\label{eq:append:approx}
&-\log p(\barS;\ \hbR,\lambda)=\nonumber\\
&-\log  p(\hat{\barS};\ \hbR,\lambda)-
\suml_k\log w(\theta^f_k,\calI^f_k)-\nonumber\\
&\frac{1}{2}\sum_k (\Delta\theta_k)^2G_k+(\Delta\calI_k)^2H_k
\end{eqnarray} 
where
\begin{eqnarray}
&G_k=-\frac{\partial^2}{\partial\theta_k^2}\tr\left[\hbR\left(\sigma^2\bI+\suml_k\calI_k\ba(\theta_k)\ba^H(\theta_k)\right)^{-1}\right]
\nonumber\\
&H_k=-\frac{\partial^2}{\partial\calI_k^2}\tr\left[\hbR\left(\sigma^2\bI+\suml_k\calI_k\ba(\theta_k)\ba^H(\theta_k)\right)^{-1}\right]
\end{eqnarray}
This implies that $\Delta\theta_k\sim \calN(0, G_k^{-1})$ and $\Delta\calI_k\sim\calN(0, H_k^{-1})$. 

\section{Perturbative KL-based Projection}
\label{appendix:perturb}
Suppose that the distribution $p(\barS_{t+1}\mid X^{(t)})$ is calculated as
\begin{equation}
p(\barS_{t+1}=\barS\mid X^{(t)})\approx p(\barS_{t}=\barS\mid X^{(t)})+\Delta p(\barS)= p(\barS;\ \hbR^+_t,\lambda^+_t)+\Delta p(\barS)
\end{equation}
The question of interest is to find the perturbation in parameters minimizing the KL distance between $p(\barS_{t+1}=\barS\mid X^{(t)})$ and the parametric model, i.e to solve
\begin{eqnarray}
&\arg\minl_{\Delta\hbR,\Delta\lambda}
-\intl_{\calS}\left(  
p(\barS;\ \hbR^+_t,\lambda^+_t)+ \Delta p(\barS)
\right)\nonumber\\
&\log\left(p(\barS;\ \hbR^+_t+\Delta\hbR,\lambda^+_t+\Delta\lambda)\right)\td\barS
\end{eqnarray}
Although this can be generally solved up to the first order, by the technique explained below, we restrict $\Delta\hbR$ to the be $\gamma\hbR^+_t$ for $\gamma>0$ to simplify calculations, and also to ensure positive semi-definiteness. After Taylor expansion, and performing the minimization, we obtain that
\begin{equation}
\gamma=-\frac{\intl_\calS\left.\frac{ \partial \log p(\barS;\ (1+\gamma)\hbR^+_t,\lambda^+_t)}{\partial\gamma}\mid_{\gamma=0}\right.\Delta p(\barS)\td\barS}{\intl_\calS p(\barS;\ \hbR^+_t,\lambda^+_t)\left.\frac{\partial^2\log p(\barS;\ (1+\gamma)\hbR^+_t,\lambda^+_t)}{\partial\gamma^2}\mid_{\gamma=0}\right. \td\barS}
\end{equation}
and
\begin{equation}
\Delta\lambda(\theta)=-\frac{\intl_\calS\frac{ \partial \log p(\barS;\ \hbR^+_t,\lambda^+_t)}{\partial\lambda(\theta)}\Delta p(\barS)\td\barS}{\intl_\calS p(\barS;\ \hbR^+_t,\lambda^+_t)\frac{\partial^2\log p(\barS;\ \hbR^+_t,\lambda^+_t)}{\partial\lambda(\theta)^2} \td\barS}
\end{equation}
We obtain the desired update by the above relations.  We further simplify this relation in favor of low complexity. Using the approximation in \eqref{eq:append:approx} and after straightforward manipulations, we get that
\begin{equation}
\gamma=\frac{\suml_k\frac{\frac{\partial G_k}{\partial\gamma}}{G_k^2}\Delta G_k+\suml_k\frac{\frac{\partial H_k}{\partial\gamma}}{H_k^2}\Delta H_k+\intl_{\Theta\times\bbR_+}\frac{\partial\log\omega}{\partial\gamma}\Delta\omega\td\theta\td\calI}
{\suml_k\frac{\left(\frac{\partial G_k}{\partial\gamma}\right)^2}{G_k^2}+\suml_k\frac{\left(\frac{\partial H_k}{\partial\gamma}\right)^2}{H_k^2}+\intl_{\Theta\times\bbR_+}\left(\frac{\partial\log\omega}{\partial\gamma}\right)^2\omega\td\theta\td\calI}
\end{equation}
and
\begin{equation}\label{eq:append:lambda}
\Delta\lambda(\theta)=\frac{\intl_{\bbR_+}\frac{\partial\log\omega}{\partial\lambda(\theta)}\Delta\omega\td\calI}
{\intl_{\bbR_+}\left(\frac{\partial\log\omega}{\partial\lambda(\theta)}\right)^2\omega\td\calI}
\end{equation}
This can be further simplified noting that the terms $\log\omega$, $G_k$ and $H_k$ are linear in $1+\gamma$ thus their partial derivative with respect to $\gamma$ equals their value at $\gamma=0$, leading to
\begin{equation}
\gamma=\frac{\suml_k\frac{\Delta G_k}{G_k}+\suml_k\frac{\Delta H_k}{H_k}+\intl_{\Theta\times\bbR_+}\log\omega\times\Delta\omega\td\theta\td\calI}
{2n+\intl_{\Theta\times\bbR_+}\left(\log\omega\right)^2\omega\td\theta\td\calI}
\end{equation}
 According to the empirical observation that the terms involving $\omega$ are substantially smaller than the other terms, we simplify the calculations more by neglecting them to obtain
\begin{equation}
\gamma=\frac{\suml_k\frac{\Delta G_k}{G_k}+\suml_k\frac{\Delta H_k}{H_k}}{2n}
\end{equation}
The expression in \eqref{eq:append:lambda}  can also be simplified by considering that $\Delta\omega\approx \delta$, and approximating $\omega$ as
\begin{equation}
\omega(\theta,\calI)\approx e^{-\calI(\lambda(\theta)-\ba^H(\theta)\bR_+^{-1}(t)\hbR_t^+\bR_+^{-1}(t)\ba^H(\theta))}
\end{equation}
Simple calculations lead to
\begin{equation}
\Delta\lambda=-\frac{1}{2}(\lambda(\theta)-\ba^H(\theta)\bR_+^{-1}(t)\hbR_t^+\bR_+^{-1}(t)\ba^H(\theta))
\intl_{\bbR_+}\calI\delta(\theta,\calI)\td\calI
\end{equation}



\ifCLASSOPTIONcaptionsoff
  \newpage
\fi



\bibliographystyle{IEEEtran}
\bibliography{IEEEabrv,bibFile}
\end{document}